\documentclass[11pt,pra,aps]{revtex4-2}

\def\by#1{#1,}
\def\and{and }
\def\yr#1{{(#1)}}
\def\paper#1{#1}
\def\jour#1{{\it #1}}

\def\vol#1{{\bf #1},}
\def\issue#1{}
\def\pages#1{\hbox{#1},}

\usepackage{amsmath}
\usepackage{amsfonts}
\usepackage{amssymb}
\usepackage{graphicx}
\usepackage{appendix}
\usepackage{placeins}

\newcommand{\pdiff}[3][]{\dfrac{\partial^{#1} #2}{\partial {#3}^{#1}}}

\begin{document}

\title{Oil slicks dynamics in semi-confined geometries in the presence of water waves: analytical and numerical insights.}

\author{Hanan Hozan}
\affiliation{School of Mathematical and Physical Sciences, University of Reading, Reading, RG6 6AX, UK}
\affiliation{Department of Mathematics, Jazan University, Jazan, Saudi Arabia}

\author{Mike Baines}
\affiliation{School of Mathematical and Physical Sciences, University of Reading, Reading, RG6 6AX, UK}

\author{Alex Lukyanov}
\email{corresponding author,\\ a.lukyanov@reading.ac.uk}
\affiliation{School of Mathematical and Physical Sciences, University of Reading, Reading, RG6 6AX, UK}

\begin{abstract}
We have shown previously that the dynamics of isolated oil slicks on the water's surface after spills is significantly influenced by surface-wave motion practically at the onset of the spreading process. In this work, we draw our attention to another practical scenario of the oil slick's behaviour in semi-confined geometries under the action of surface waves and in the presence of oil leakage source. We hypothesize that the geometric constraints should qualitatively change the oil layer response to the wave motion leading to localization of oil spill domains even at modest levels of the surface-wave perturbations. Several realistic cases were rigorously explored, with special attention paid to the interplay, the combined effect of the two factors, oil influx and wave motion. It was demonstrated quantitatively how the spreading process can be either facilitated or suppressed leading to potential confinement.  
\end{abstract}

\maketitle
 
\section{Introduction}

The dynamics of oil spills on the sea surface under realistic environmental conditions represents a complex phenomenon with practical applications~\cite{GasIndustry2016, Garcia2020, Barron2020, Boufadel2023}. The spreading events are influenced by a variety of natural factors such as winds, currents and water waves leading to oil evaporation and emulsification by breaking waves with the subsequent dispersion due to turbulent motion. The result of these diverse environmental conditions is that the oil spill domains in the long run contain both continuous surface layers of oil and isolated oil droplets forming emulsive fraction in the bulk water column~\cite{VMHD2017}. It is the emulsified oil phase that presents decontamination challenges as the droplets are easily entrained to the water column to different depths by various turbulence conditions. Therefore, the focus of a substantial body of research work was specifically on the emulsification processes, and the distribution and dynamics of oil droplets in the marine environment resulted from oil spills~\cite{Delvigne1988, Li2007, Li2008, Fingas2011, Reed2015, Li2017, Katz2017, Zhao2022}. This information is essential for oil spreading modelling used for emergency planning and environmental impact assessments~\cite{SP2017, ReviewSpill2021, Zhao2014}.

The current conceptual frameworks for the dispersion of oil in the water column still relies on the seminal work of Delvigne and Sweeney (1988), who conducted breaking wave experiments in wave tanks and measured the oil droplet distribution and the oil mass in the water column after breaking events~\cite{Delvigne1988}. These experiments, analysis and further studies helped to formulate empirical relationships between the oil entrainment and the problem parameters such as oil viscosity, temperature and the intensity of the breaking wave. While it was also found that the entrainment amount of oil is practically independent of the oil layer thickness during a particular breaking event, the net result, the ultimate distribution is conditioned by the oil layer distribution obtained on the time scale  of the gravity-viscous characteristic spreading time, which is of several hours~\cite{Fay1969, Fay1971, Hoult1972, Katz2017}. 

The main objective of the current study is to understand the formation of that preconditioning oil layer distribution rigorously taking into account the effect of the surface wave motion. We have already shown previously on the basis of a thin film model and the linear deep-water wave approximation that a single harmonic of the surface wave depending on the nature (a travelling wave or a standing wave) can lead in general to a drift and depletion of an oil slick. In the current work, we consider some important special cases, which require particular attention, namely the evolution of an oil slick in the presence of a source of the oil on wavy water surfaces in semi-confined surface geometries. The problem formulation allows for a control of the oil spreading, and this will be interesting to investigate these cases using the previously developed thin film model. 

\begin{figure}[ht!]
\begin{center}
\includegraphics[trim=0.3cm 2.cm 1cm -0.5cm,width=0.7\columnwidth]{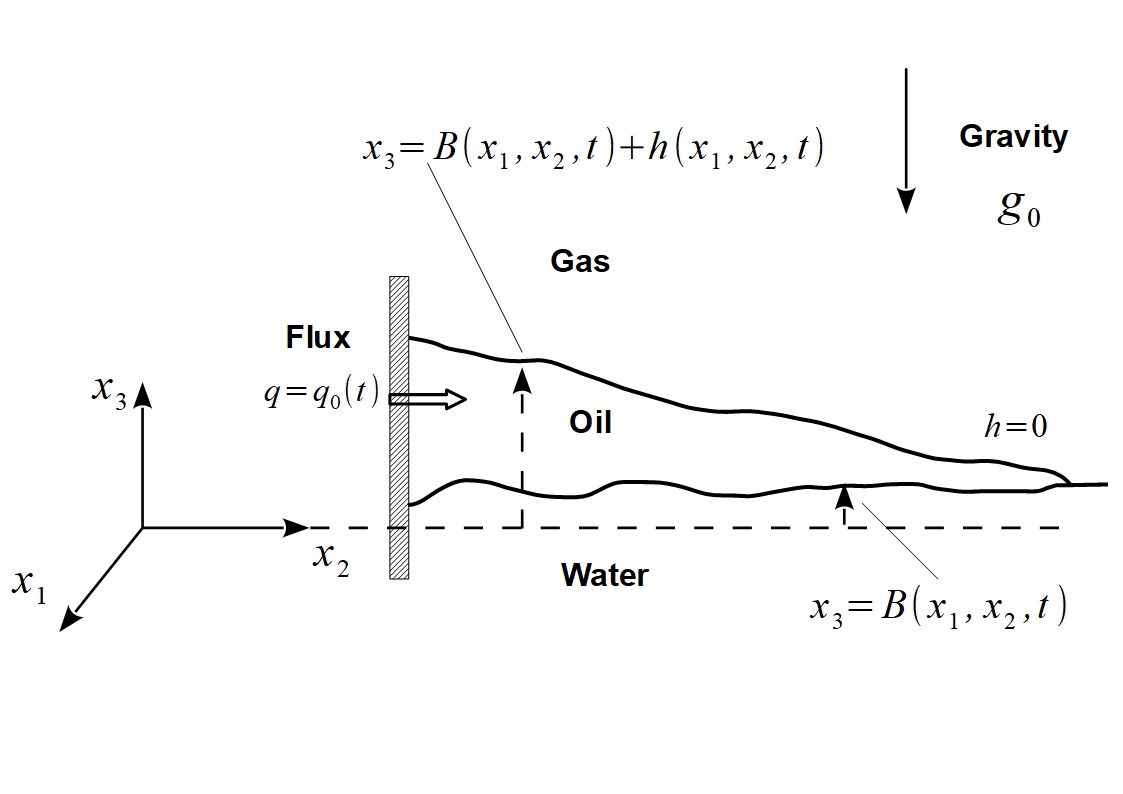}
\end{center}
\caption{Illustration of the oil layer on the water surface problem geometry.} 
\label{Fig1}
\end{figure}

In the subsequent sections, we very briefly introduce the mathematical model based on the thin film approximation. The model then is analyzed analytically to obtain some reference self-similar solutions in the presence of the oil source and numerically to investigate various dynamics of the oil slick. First, we study the effects brought into the oil slick dynamics due to the presence of a source in the absence of the water surface disturbances. Then we carefully analyse the variations in the dynamics induced by the water wave motion.

\section{Problem formulation and the mathematical model}

Consider an oil layer with density $\rho$ and dynamic viscosity $\mu$, spreading on the water's wavy surface, as depicted in Fig. \ref{Fig1}. In this study, we assume the oil is Newtonian, although it's noted that various crude oil compositions could exhibit non-Newtonian behavior.

The positions of the oil-water and oil-gas interfaces are denoted by $x_3=B(x_1,x_2,t)$ and $x_3=B(x_1,x_2,t) + h(x_1,x_2,t)$, respectively, where $h$ represents the thickness of the oil layer.

The mathematical framework is identical to the previous study~\cite{Hazan2024}. In the thin-film approximation, the model distinguishes between two length scales: $L$ in the $x_{1,2}$-directions, which might represent the diameter of the oil slick or the wavelength of surface disturbances, and $H$ in the $x_3$-direction, corresponding to the oil layer's thickness~\cite{Matar2009}.

The ratio of these scales, defined as $\frac{H}{L} = \varepsilon$, is considered a small parameter, $\varepsilon \ll 1$. This assumption is based on the typical ranges of oil-slick characteristic lengths, with $1\,\text{m} \leq L \leq 100\,\text{m}$ covering either the oil slick's size or the scale of perturbations, and $10^{-4}\,\text{m} \leq H \leq 10^{-2}\,\text{m}$ representing the thickness of the oil layer.

The governing equations in the thin film approximation are obtained by using the small parameter $\varepsilon$ and averaging the Navier-Stokes equations over the oil layer, that is by introducing the liquid fluxes 
$$
q_{1,2} = \int_B^{h+B}\, v_{1,2}\, dx_3,
$$
where $v_{1,2}$ are the components of the velocity vector, and using boundary conditions on the interfaces. 

In what follows, the mathematical formulation is brought into a non-dimensional form (unless specified otherwise) by introducing  
$$
\widetilde{x}_{1,2}=x_{1,2}/L,\quad\widetilde{x}_3=x_3/H=x_3/\varepsilon L,
$$
$$
\widetilde{v}_{1,2}=v_{1,2}/U,\quad\widetilde{v}_3= v_3/\varepsilon U,\quad\widetilde{t}=t/t_0,
$$ 
where $U$ is the characteristic velocity and $t_0=L/U$. The pressure in the system is assumed to be dominated by viscous contributions, so that it was normalised by $p_0=\frac{\mu U}{\varepsilon H}$, $\tilde{p}=p/p_0$.

In the thin film approximation, neglecting surface tension effects, one gets
\begin{equation}
\label{LE2}
\frac{\partial h}{\partial t}  + \frac{\partial q_1}{\partial x_1} + \frac{\partial q_2}{\partial x_2} = 0,
\end{equation}
$$
\frac{\partial q_1}{\partial t} +  \frac{\partial }{\partial x_1} \left(\frac{6}{5} \frac{q_1^2}{h} -\frac{2}{5}(q_1 V_1) + \frac{1}{5}V_1^2 h \right) +
$$
\begin{equation}
\label{GE2D3s}
 \frac{\partial }{\partial x_2} \left( \frac{6}{5} \frac{q_1q_2}{h} -\frac{1}{5}(q_1 V_2 + q_2 V_1) + \frac{1}{5} V_1 V_2 h \right) = 
\end{equation}
$$
- Fr^{-2}\, h\, \pdiff{(h+B)}{x_1} -\frac{3}{Re}\frac{q_{1}-V_1 h}{h^2}
$$
and
$$
 \frac{\partial q_2}{\partial t} + \frac{\partial }{\partial x_1} \left( \frac{6}{5} \frac{q_1q_2}{h} -\frac{1}{5}(q_1 V_2 + q_2 V_1) + \frac{1}{5} V_1 V_2 h + \right) +
$$
\begin{equation}
\label{GE2D4s}
 \frac{\partial }{\partial x_2} \left( \frac{6}{5} \frac{q_2^2}{h} -\frac{2}{5}(q_2 V_2) + \frac{1}{5} V_2^2 h\right)=
\end{equation}

$$
- Fr^{-2}\, h\,  \pdiff{(h+B)}{x_2}  -\frac{3}{Re}\frac{q_{2}-V_2 h}{h^2}.
$$
Here, $Re=\varepsilon \frac{\rho U H}{\mu}$ and $Fr^2=\frac{U^2}{g_0 H}$ are the non-dimensional Reynolds and Froude numbers respectively.

In the limit of small Reynolds number $Re\ll 1$, the system of equations is reduced to a single non-linear advection-diffusion equation, which is in the one-dimensional case

\begin{equation}
\label{LE2TH1D}
 \frac{\partial h}{\partial t}  +  \frac{\partial q}{\partial x}  = 0,
\end{equation}
$$
q = -  \frac{\alpha_g}{3}\pdiff{(h+B)}{x} h^3 +V h,
$$
where parameter $\alpha_g = \frac{\rho g_0 \varepsilon H^2}{\mu U}$ is representing a non-dimensional measure of gravity to viscous forces. Apparently, this parameter also characterises the strength of the effective diffusivity, caused by the competition of gravity and viscous forces in the spreading process, in comparison to the advection in the model. Obviously, the larger $\alpha_g$, the larger is the contribution of diffusion.

To analyse the effects of the external forcing due to the oil influx and the wave motion, we consider a problem posed on a compact domain $x_l\le x \le x_r$.

At one end, which is free to move, one has 
\begin{equation}
\label{BVLR}
h(x_r) = 0
\end{equation}
and by conservation of mass there exists a boundary velocity
\begin{equation}
\label{BVR}
\frac{d x_r}{dt} =\left. \frac{q}{h}\right|_{x=x_r} =  \left.\left(-\frac{\alpha_g}{3}\pdiff{(h+B)}{x} h^2 +V\right)\right|_{x=x_r}.
\end{equation}

The other end, which is conveniently set at $x=x_l=0$, where the source of oil is located, is fixed.  As the boundary condition at this end, the influx rate is given as a function of time
\begin{equation}
\label{BFL}
q(0)= \left.\left(-\frac{\alpha_g}{3}\pdiff{(h+B)}{x} h^3 + V h\right)\right|_{x=0} =q_0(t).
\end{equation}

The water-oil interface perturbations, that is functions $B$ and $V$, are taken, as in our previous study~\cite{Hazan2024}, still as harmonics of the deep-water waves in the linear approximation 
\begin{equation}
\label{DPWTH}
B(x,t) = \sum_j B_0^j\sin(k_j x - \omega_j t), 
\end{equation}
$$
V(x,t) =\sum_j  B_0^j \varepsilon \omega_j \sin(k_j x - \omega_j t).
$$ 

Here $B_0^j$ is the amplitude, $\omega_j$ is the frequency and $k_j$ is the wave vector. In non-dimensional form, the dispersion relation is
\begin{equation}
\label{DispR}
\omega_j^2 = \varepsilon^{-1}\, Fr^{-2}\, k_j.
\end{equation}

The ratio of characteristic amplitude of the velocity perturbations $V_0=B_0 \varepsilon \omega$ to the rate of diffusion, $Pe_E=\frac{V_0}{\alpha_g}$ can be regarded as an effective P\'{e}clet  number.

\section{Spreading in semi-confined geometry}

At first, we consider a situation with a source in the absence of perturbations. This special case should help us to understand the effects introduced by the external source and their parametric dependencies. 

\subsection{Self-similar behavior in the presence of an influx source}

The governing equation, disregarding surface waves, is reduced to a non-linear diffusion equation of the form, 
\begin{equation}
\label{NND1D}
 \frac{\partial h}{\partial t}  =  A \frac{\partial }{\partial x} \left( h^m \pdiff{h}{x}\right),
\end{equation}
where $A=\frac{\alpha_g}{3}$ and $m=3$.

The problem is augmented with the set of boundary conditions (\ref{BVLR})-(\ref{BFL}) corresponding to the external source at $x=0$, which takes the form
\begin{equation}
\label{BVLRE}
h(x_r) =0,
\end{equation} 
\begin{equation}
\label{BVRDE}
\frac{d x_r}{dt} = -  A\left. \pdiff{h}{x} h^{m-1}\right|_{x=x_r}
\end{equation}
and at $x=0$
\begin{equation}
\label{BFLE}
A\left. \pdiff{h}{x} h^{m}\right|_{x=0}=q_0(t).
\end{equation}

As is well known, equation (\ref{NND1D}) is invariant with respect to transformation of variables $t^{\prime}=t/\eta,\, x^{\prime}=x/\eta^{\beta}$ and $h^{\prime}=h\eta^{\alpha}$, for any positive constant $\eta>0$, if $m\alpha + 2\beta = 1$. In self-similar variables $w=x\, t^{-\beta}$ and $\phi(w)=ht^{\alpha}$, the governing equation (\ref{NND1D}) takes the general form
 \begin{equation}
		\label{SelfSGE}
		\beta w \frac{d \phi}{d w} + \alpha \phi+A \frac{d}{dw}\left ( \phi^{m} \frac{d\phi}{dw}\right)=0.
	\end{equation}

The boundary value problem in the case of an isolated oil spot, that is with both free ends subject to the boundary conditions (\ref{BVLRE})-(\ref{BVRDE}) where $h=0$ and no external source present, is also invariant to the transformation of variables. If we consider symmetric solutions, problem (\ref{SelfSGE}) can be augmented with a set of boundary conditions
	\begin{equation}
		\label{SymmBCSSV}
		\left. \phi({w}) \right|_{w= \hat{w}}=0,\quad \left. \frac{d\phi}{dw}\right|_{w=0} = 0,
\end{equation}
where due to conservation of mass, $\alpha=\beta=\beta_0=\frac{1}{m+2}$ and parameter $\hat{w}$ is defined by the total amount of the liquid and the power of the non-linearity. That is
	\begin{equation}
		\label{constw}
		\hat{w}=\left[\left(\frac{A(2+m)}{6}\right)^{1/m} \left(\frac{\Gamma\left(\frac{2(1+m)}{m}\right)} {\left(\Gamma\left(\frac{1+m}{m}\right)\right)^2} \right) I_0\right]^{\frac{m}{2+m}}.
	\end{equation}
Here, $\Gamma$ is the Gamma function and $I_0=2\int_{0}^{x_r}\, h\, dx$ is the total initial amount of the oil.

While it is difficult to obtain a closed-form solution to problem (\ref{SelfSGE}) in general, it can be integrated exactly at $\alpha=\beta=\beta_0=\frac{1}{m+2}$ to obtain, after applying the boundary conditions (\ref{SymmBCSSV}), a self-similar solution
	\begin{equation}
		\label{SSS}
		\phi(w)= \left(\frac{m}{2A(m+2)}\right)^{1/m} \left(  \hat{w}^{2}- w^{2}\right)^{1/m}.
	\end{equation}

The self-similar solution (\ref{SSS}) is asymptotic, so that arbitrary initial data evolve into it in time. Therefore, in the limit $t\to\infty$, one can expect, as one can see from the solution, that the domain boundary is moving as 
\begin{equation}
\label{ScB}
x_{r}(t)\propto A^{\frac{1}{2+m}}\, I_0^{\frac{m}{2+m}}\, t^{\beta_0},
\end{equation}  
while the maximum value of the oil slick height $\max(h)\propto t^{-\beta_0}$.

That is, as one would naturally expect in the absence of an oil source, the initial liquid volume is spreading out, while the height is decreasing. 

The problem with the external source present is not expected to demonstrate the same self-similar properties, as the boundary condition at $x=0$ may not satisfy the invariant properties in general.

We consider a special case 
$$
q_0=Q_0 t^{\gamma}
$$
where self-similarity could still be observed. Here, $Q_0$ is some constant parameter. Then, one can easily demonstrate that the self-similar behaviour should be observed if 
$$
\alpha=-\frac{2\gamma + 1}{m+2}, \quad \beta = \frac{1+m(\gamma+1)}{m+2}.
$$

In this case, problem (\ref{SelfSGE}) can be augmented with 
 \begin{equation}
		\label{BCF}
		\left.
		\phi({w})
		\right|_{w=\hat{w}}=0, \quad \left. -A \phi^m \pdiff{\phi}{w} \right|_{w=0} = Q_0,
\end{equation}
where $\hat{w}$ can be found from the amount at a given moment of time.

While it is difficult to obtain a closed-form solution to (\ref{SelfSGE}) subject to boundary conditions (\ref{BCF}), it is not that difficult to see that the front motion is expected to follow $x_r(t)\propto t^{\beta}$,	 which is quite different in general from the asymptotic behaviour in the isolated spot case. In particular, at $\gamma=0$ and $m=3$, when the external source is independent of time, $\beta=\frac{m+1}{m+2}=\frac{4}{5}>\beta_0$, the difference is dramatic, as one can observe much larger exponent, which is, we note, independent of the external flux strength. 

We also note, that in the special case $\gamma=-1$, the exponents are $\alpha=\beta=\beta_0$, so that the dynamics is expected to be very close to the isolated spot case, but not identical, as the problem is subject to different boundary conditions (\ref{BCF}).

\begin{figure}[ht!]
\begin{center}
\includegraphics[trim=0.3cm 2.cm 1cm -0.5cm,width=0.7\columnwidth]{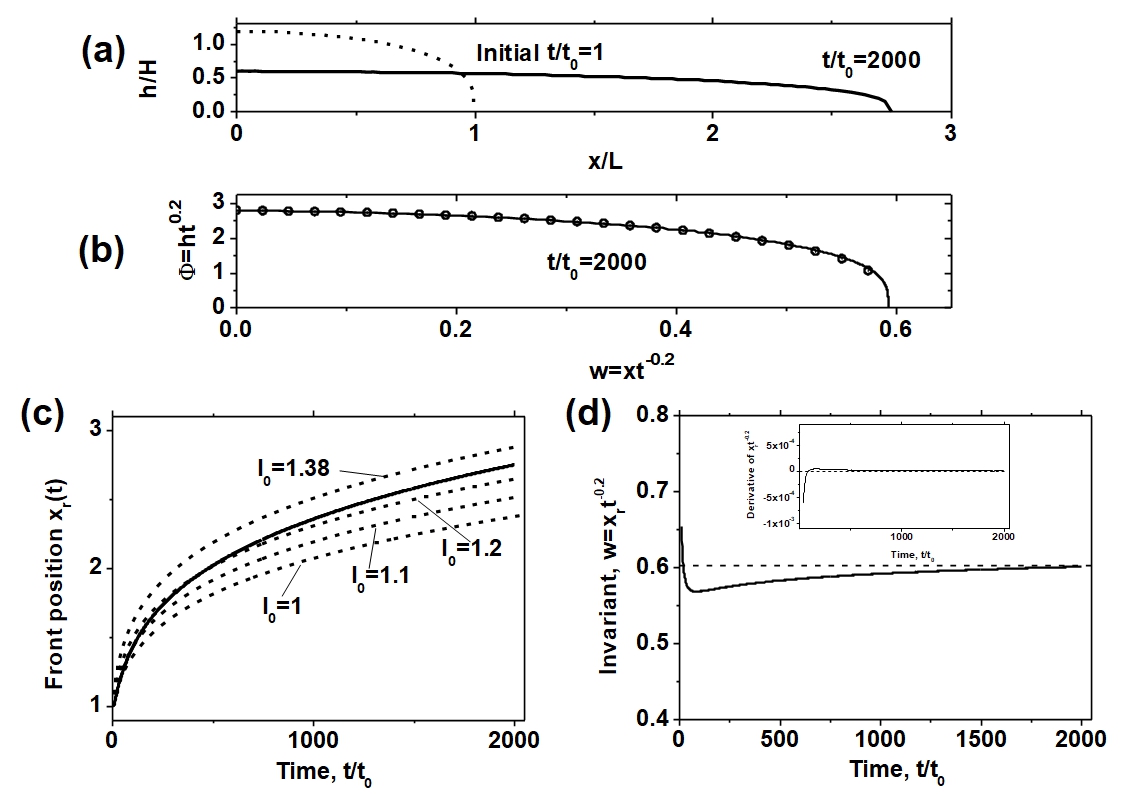}
\end{center}
\caption{Illustration of the oil layer dynamics in the presence of external source $q(t)=Q_0 t^{-1}$, $Q_0=0.05$ and in the absence of the wave motion at $\alpha_g=2\cdot 10^{-2}$ and $Fr^2=10$. (a) Evolution of the initial profile till $t/t_0=2000$. The initial profile at $t/t_0=1$ was the self-similar solution (\ref{SSS}) at $I_0=1$. (b) Evolution of the profile in terms of self-similar variables $\phi=ht^{0.2}$ and $w=xt^{-0.2}$. The symbols are for the solution of (\ref{SelfSGE}) with (\ref{BCF}) obtained numerically. (c) Evolution of the moving front $x_r(t)$ initially allocated at $x=1$ at $Q_0=0.05$ and $I_0=1$. The dashed lines designate the reference solutions at $Q_0=0$ at different initial amounts $I_0$.  (d) Evolution of the invariant combination $x_r t^{-0.2}$ at $Q_0=0.05$ and $I_0=1$ in self-similar variables. The insert shows its time derivative.} 
\label{Fig2}
\end{figure}
	
The specific asymptotic behaviour in the special case $\gamma=-1$ is apparently due to the vanishing external source strength at $t\to\infty$, therefore leading to the similar behaviour as in the case of no external flux present.

\subsubsection*{Self-similar behaviour in a radially symmetric case.}

It is worth considering briefly a special radially symmetric case. The problem is only slightly modified preserving its generic conservative form in the case of a radially symmetric solution. That is
\begin{equation}
\label{LE2TH1DRS}
 \frac{\partial h}{\partial t}  +  \frac{1}{r}\frac{\partial( qr)}{\partial r}  = 0, \quad q = -  \frac{\alpha_g}{3}\pdiff{h}{r} h^3.
\end{equation}

At the moving boundary of a closed domain $R=R(t)$, 
$$
h(R)=0
$$
and
$$
\frac{dR}{dt} = \frac{q}{h} = -\frac{\alpha_g}{3}\pdiff{h}{r} h^2.
$$

In the absence of external flux, at the centre,
$$
\frac{\partial h}{\partial r} = 0.
$$

It is then not difficult to show that in that one-dimensional, radially symmetric case, the scaling will be quite similar. That is, assuming, as before, transformation of variables $t^{\prime}=t/\eta,\, r^{\prime}=r/\eta^{\beta_r}$ and $h^{\prime}=h\eta^{\alpha_r}$, and the conservation of the total amount $\int _0^R \, h\, r dr=const$, the parameters of the transformation are
$$
\beta_r^0 = \frac{1}{2+2m}, \quad \alpha_r^0 = \frac{2}{2+2m},
$$
suggesting that the combination $\frac{r}{t^{\beta^0_r}}$ is invariant and hence 
$$
R(t) \propto t^{\beta^0_r}.
$$

In general, if the external flux is present, this is equivalent to state that at some $r_0\ll R(t)$,
$$
2\pi q r_0 = Q_0 t^{\gamma}.
$$
As a result
$$
\beta_r = \frac{1 + m(\gamma+1)}{2+2m}, \quad \alpha_r = -\frac{2\gamma}{2+2m}
$$
and 
$$R(t)\propto t^{\beta_r}.$$ Considering particular values of $\gamma>0$ and $m=3$, one can observe that $\beta_r> \beta_r^0$.

\subsection{Numerical tests and benchmarking}

\begin{figure}[ht!]
\begin{center}
\includegraphics[trim=0.3cm 2.cm 1cm -0.5cm,width=0.7\columnwidth]{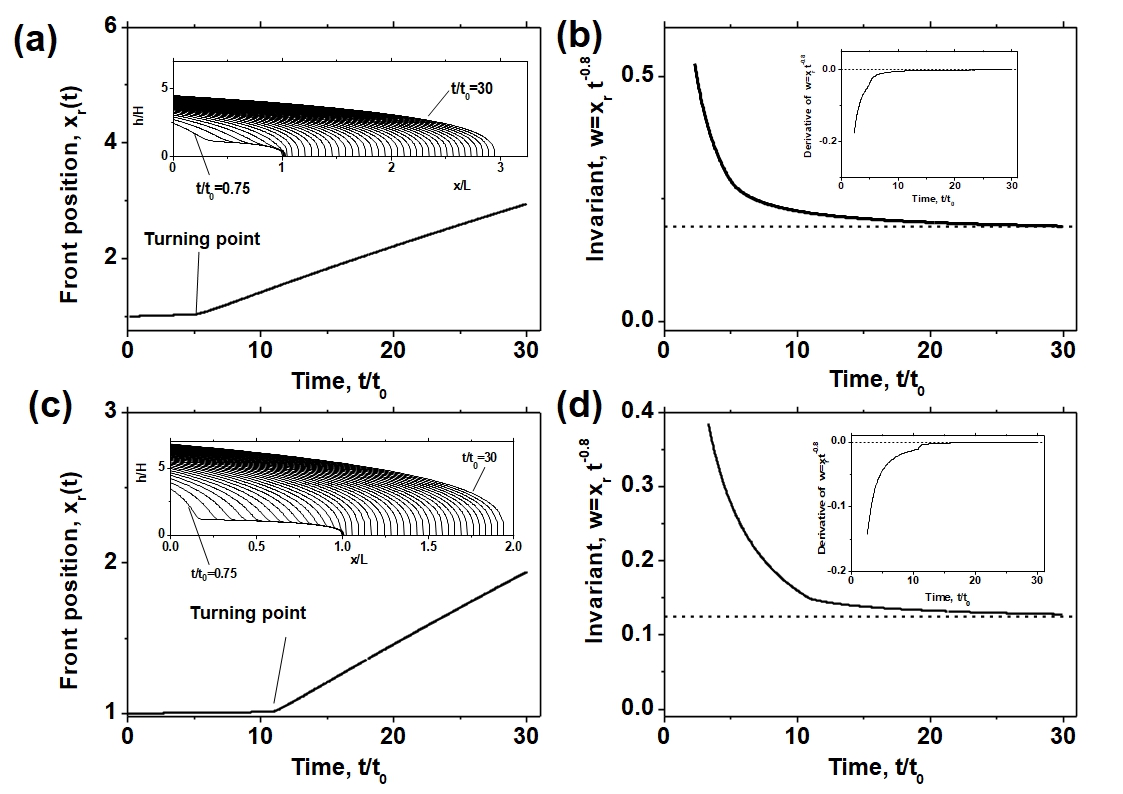}
\end{center}
\caption{Illustration of the oil layer dynamics in the presence of external source $q(t)=Q_0$ and in the absence of the wave motion. (a) Evolution of the profile $h(x,t)$ and the moving front at $\alpha_g=2\cdot 10^{-2}$, $I_0=1$, $Fr^2=10$ and $Q_0=0.3$. (b) Evolution of the moving front $x_r t^{-0.8}$ at $\alpha_g=2\cdot 10^{-2}$, $I_0=1$, $Fr^2=10$ and $Q_0=0.3$ in terms of self-similar variables. The insert shows its time derivative. (c) Evolution of the profile $h(x,t)$ and the moving front $x_r(t)$  at $\alpha_g=0.0025$, $I_0=1$, $Fr^2=20$ and $Q_0=0.3$.  (d) Evolution of the moving front $x_r t^{-0.8}$ at $\alpha_g=0.0025$, $I_0=1$, $Fr^2=20$ and $Q_0=0.3$ in terms of self-similar variables. The insert shows its time derivative.} 
\label{Fig3}
\end{figure}

\begin{figure}[ht!]
\begin{center}
\includegraphics[trim=0.3cm 0.3cm 1cm -0.5cm,width=0.6\columnwidth]{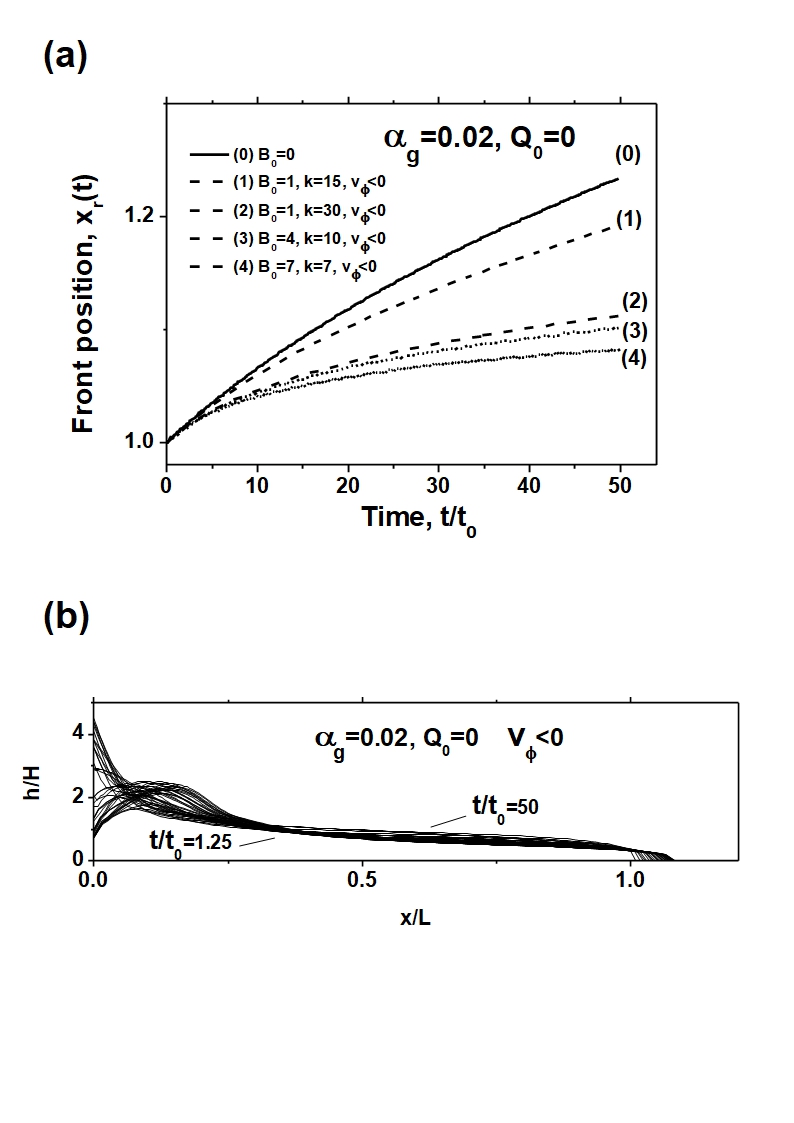}
\end{center}
\caption{Illustration of the oil layer dynamics in the presence of the wave motion at $v_{\phi}<0$ in semi-confined geometry at $Q_0=0$, $I_0=1$, $\alpha_g=0.02$ and $Fr^2=10$. (a) Evolution of the moving front $x_r(t)$ as a function of time at different $B_0, k,\omega$. (b) Evolution of the oil layer profiles at $B_0=7$, $k=7$ and $\omega\approx 60$.} 
\label{Fig5}
\end{figure}

\begin{figure}[ht!]
\begin{center}
\includegraphics[trim=0.3cm 2.cm 1cm -0.5cm,width=0.7\columnwidth]{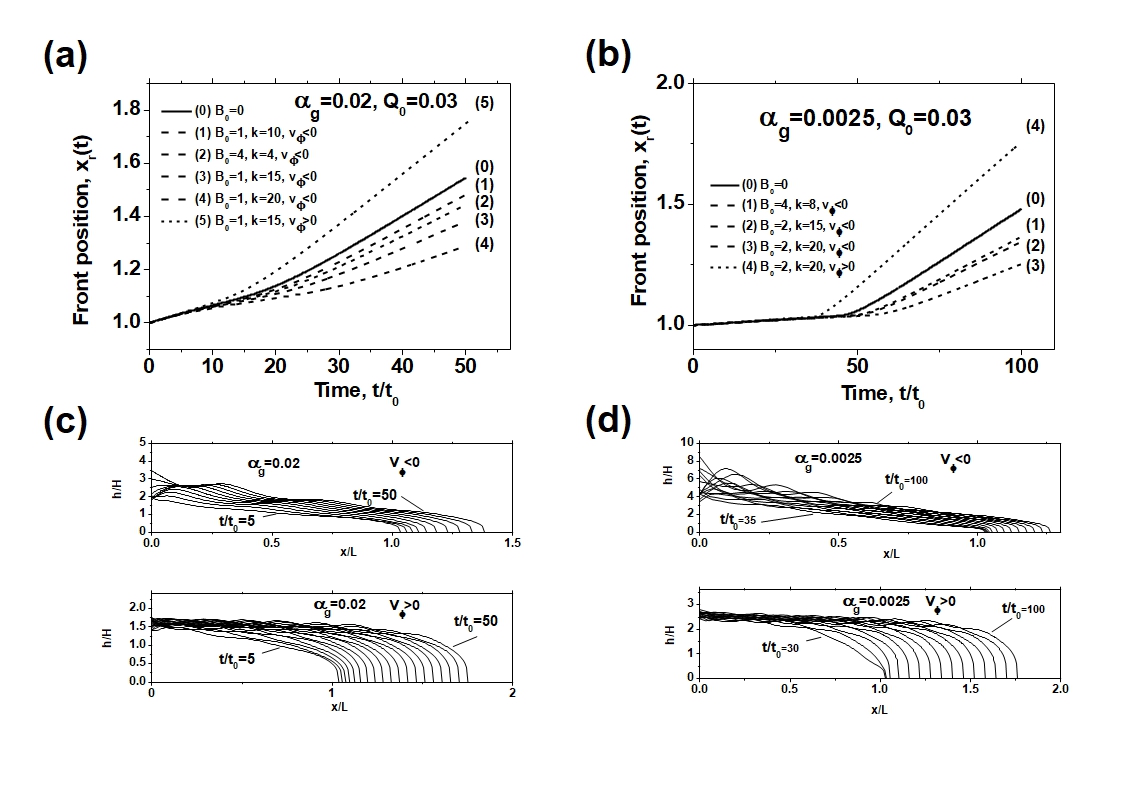}
\end{center}
\caption{Illustration of the oil layer dynamics in the presence of external source at $\gamma=0$ ($q(t)=Q_0$, $Q_0=0.03$) and the wave motion when initially $I_0=1$. (a) Evolution of the moving front $x_r(t)$ as a function of time at $\alpha_g=2\cdot 10^{-2}$ and $Fr^2=10$ at different $B_0, \omega$ and $k$. (b) Evolution of the moving front $x_r(t)$ as a function of time at $\alpha_g=2.5\cdot 10^{-3}$ and $Fr^2=20$ at different $B_0, \omega$ and $k$.  (c) Evolution of the oil layer profiles at either $v_{\phi}<0$ or $v_{\phi}>0$ and at $\alpha_g=0.02$, $Fr^2=10$, $B_0=1$, $k=15$ and $\omega\approx 87$.  (d) Evolution of the profiles at either $v_{\phi}<0$ or $v_{\phi}>0$ and at $\alpha_g=0.0025$, $Fr^2=20$, $B_0=2$, $k=20$ and $\omega = 100$.} 
\label{Fig4}
\end{figure}

To verify the self-similar behaviour in the presence of external flux, we consider numerical solutions of the problem. The numerical simulations are performed using the moving-mesh method~\cite{Baines2015} with some adjustments to account for the external oil source. In particular, in the numerical method, the mesh points are moving with the Lagrangian velocity similar to that of the moving boundary (\ref{BVR}) but defined at the mesh points. Therefore, the total amount of the oil in each simulation interval between the mesh points is conserved in the absence of the source term. When the external source at $x=0$ is present, the conservation principle results in the first interval becoming larger and larger therefore reducing the accuracy of the scheme. To mitigate the effect of the disproportionate enhancement of the first interval, new mesh points are added in the first interval to keep the accuracy on the same level as that at the beginning of the simulations.

In the simulations, initially, the oil layer domain is conveniently set to be $x\in[0, 1]$ with the total initial amount of the oil $I_0=\int_{0}^{1}\, u\, dx = 1$, unless otherwise specified, to obtain initial layer thickness of the order of one. As we will see later, the existence of a fixed, that is not moving boundary makes the water-wave effect much stronger allowing much lower wave amplitudes to observe the effect.

We mostly consider a particular set of dimensional and non-dimensional parameters corresponding to the initial phases of the oil spill dynamics. The oil layer thickness is taken in between $H=10\,\mbox{mm}$ and $H=1\,\mbox{mm}$. The horizontal length scale and the characteristic velocity are taken at $L=50\,\mbox{m}$ and $U=1\,\mbox{m}/\mbox{s}$ respectively, so that the characteristic time $t_0=50\,\mbox{s}$. 

If we take $\rho=8.7\cdot 10^{2}\,\mbox{kg}/\mbox{m}^3$ and $\mu=8.7\cdot 10^{-3}\,\mbox{Pa}\cdot\mbox{s}$ corresponding to light oil, the Reynolds and Froude numbers varies in between $0.002 \le Re\le 0.2$ and $100\ge Fr\ge 10$ respectively, and $2\cdot 10^{-5} \le \alpha_g \le 0.02$. 

\subsubsection*{Vanishing in time external source}

We first consider the special case when $\gamma=-1$, so that self-similarity exponents $\alpha=\beta=\beta_0=\frac{1}{5}$ are as in the isolated spot case. 

The evolutions of the initial profiles and the front are shown in Fig. \ref{Fig2}. The evolved profile in the self-similar variables corresponds well to the self-similar solutions expected from (\ref{SelfSGE}) with the boundary conditions (\ref{BCF}), Fig. \ref{Fig2} (b). 

 The evolution of the moving front, Fig. \ref{Fig2} (c), demonstrated much faster dynamics in comparison to the reference case at the same initial amount $I_0=1$ and $Q_0=0$. The effect appears to be mostly due to the variation of the total amount in the process. Indeed, in the absence of the external source, the front motion, as is suggested by (\ref{ScB}), should follow $x_r\propto I_0^{3/5}$ at a fixed moment of time. This scaling is observed as one can see from the reference solutions in panel (c) at $Q_0=0$ and at different initial amounts $I_0=1.1$, $I_0=1.2$ and $I_0=1.38$. The last value  corresponds to the total amount $\int_{0}^{x_r}\, u\, dx \approx 1.38$ achieved due to the source $q_0=Q_0 t^{-1}$ at $t/t_0=2000$ and $Q_0=0.05$. Therefore, while the dynamics is not identical to the isolated spot case, one can see from the comparison that the effect is primarily due to the increasing amount of the oil in the domain.

The behaviour of the invariant combination $x_r t^{-0.2}$, assuming, of course, self-similar trends, shown in Fig. \ref{Fig2} (d) indeed demonstrates that in the long run, the dynamics correspond well to the theoretical scaling expectations.

\subsubsection*{Constant external source}

We have further verified, that the self-similar behaviour predicted theoretically is also observed in the numerical solutions in another important case with $\gamma=0$ when $q_0=Q_0$, Fig. \ref{Fig3}. The numerical simulations have been done in a similar way. As one can observe following the potentially invariant combination in this case $x_r t^{-0.8}$, the dynamics indeed tends to the theoretically expected behaviour, Fig. \ref{Fig3} (b,d). To observe transient behaviour, the influx value was set to a sufficiently large value so that during the simulation time interval the total variation of the oil amount would substantially exceed the initial value. One can observe, Fig. \ref{Fig3} (a,c) that there is an initial slow propagation mode when the external flux has no influence on the moving front, followed by a short transition period at a turning point $t/t_0\approx 5\div 10$ to the self-similar behaviour when the front motion is driven by the external flux.

\subsection{Spreading in the presence of traveling-wave perturbations}

Now, as we have established the main characteristic features introduced by the presence of the external source, one can turn attention to the effects caused by the surface waves. We analyze a representative case of a single harmonic mode in the form of a wave travelling in either positive or negative direction with phase velocity $v_{\phi} = \pm \frac{\omega}{k}$
\begin{equation}
\label{DPWTH1}
B(x,t) =  B_0\sin(k x \pm \omega t), 
\end{equation}
$$
V(x,t) =  B_0 \varepsilon \omega \sin(k x \pm \omega t).
$$
One can anticipate, on the basis of the previous results~\cite{Hazan2024}, that the waves travelling in the direction to the restricting boundary at $x=0$  ($v_{\phi}<0$) could possibly lead to the decrease of the front velocity of the oil-occupied domain resulting in a confinement. On the other hand, the waves travelling in the opposite direction ($v_{\phi}>0$) are expected to further facilitate the spreading.

To illustrate the effect of the traveling-wave perturbations, we consider motion of the front and evolution of the profiles. The results of simulations with and without external source of the oil are shown in Fig. \ref{Fig5} (a) - (b) and \ref{Fig4} (a)-(d) at different water-wave parameters. 

Consider first the case when no external source is present, that is at $Q_0=0$. In the absence of the external source, the dynamics is defined by the initial amount and by parameter $\alpha_g$. Parameter $\alpha_g=0.02$ was chosen to maximize the diffusion effects. This corresponds to the earlier stages of the spreading event as is shown in the previous studies~\cite{Hazan2024}. As one can see, Fig. \ref{Fig5} (a) - (b), even at this value of $\alpha_g$ when the diffusion counteractive process is strong, the advance of the domain boundary can be completely inhibited by the oncoming wave ($v_{\phi}<0$) at relatively low amplitudes $B_0$. 

Note that earlier, when analyzing the case of an isolated spot~\cite{Hazan2024}, we found that the dynamics of the front is controlled, roughly, by a combination of two parameters $B_0 k$.  A similar trend is observed in the semi-confined geometry, Fig. \ref{Fig5} (a), where one can see that there is a marginal change in the overall effect when parameter $B_0 k$ varies from $30$ to $40$, while the pairs of values $B_0$ and $k$ are $(1,30)$ and $(4,10)$. Therefore, shorter wavelength perturbations should always have a more profound effect on the front motion. 

If we now consider the scenario with an external source present, one can observe that in the similar range of $B_0 k\approx 20$, the propagation can be partially suppressed, though not completely, by the oncoming waves $v_{\phi}<0$. So that indeed, the effect is similar to the previous case, but the external source makes it more difficult to suppress the propagation completely. Basically, the front motion is controlled by the external source amplitude $Q_0$, and the inhibition effect is independent of the diffusion parameter $\alpha_g$. This is seen in Fig. \ref{Fig4}. At the same time, the waves moving in the opposite direction ($v_{\phi}>0$) can clearly facilitate the spreading.  

\section{Conclusions}

We have analysed the dynamics of oil spots in a semi-confined geometry with a limiting boundary either with or without external source of oil. The presence of an external source predictably changes (increases) characteristic exponents of the spreading process and increases the boundary velocity at other parameters fixed. 

It is interesting to note that the characteristic exponents of the front motion are independent of the strength of the external source, and may serve potentially as an indicator. At the same time, the front velocity, of course, scales with the external source strength after some initial waiting time, which is required for the external source influence to propagate through the domain. The front motion is clearly affected by the oncoming water waves. It can be completely inhibited  by the oncoming water waves in the absence of the external source. The presence of the external source makes it more difficult to suppress the front motion. In the absolute terms, one can conclude, that the oil slick may become trapped in the domain by the wave motion with the amplitude on the scale (as $H=10\,\mbox{mm}$) of a few centimeters in absolute values.

At the same time, as it might be anticipated, the co-propagating waves would only enhance the front velocity leading faster spreading.

This would be interesting to consider the effects of the wave motion in two-dimensional settings and using realistic spectrum of ocean waves, for example~\cite{Moskowitz1964}. 

\bigskip

\centerline{\bf Declaration of competing interest}

The authors declare that they have no known competing financial interests or personal relationships that could have appeared to influence the work reported in this paper.

\bigskip

\centerline{\bf Author Contributions Statement}

All authors were involved in the preparation of the manuscript. All authors gave final approval for publication and agree to be held accountable for the work performed therein.

\begin{acknowledgments}
HH was supported through a PhD scholarship awarded by Jazan University, Saudi Arabia. 
\end{acknowledgments}

\end{document}